\documentclass[aps,prl,floatfix,twocolumn,amsfonts,amssymb,amsmath]{revtex4-1}

\usepackage[colorlinks=true,linkcolor=black,citecolor=black,filecolor=black,urlcolor=black,breaklinks=true]{hyperref}
\usepackage{dsfont} 
\usepackage{float} 
\usepackage{mathtools}
\usepackage[capitalize]{cleveref}
\usepackage{physics}
\usepackage{tikz}
\usepackage{braket}
\usepackage{multibib}

\newcommand{\m}{\mathrm}
\newcommand{\Ito}{It\=o\ }
\newcommand{\I}{\m{i}}
\newcommand{\cald}{\mathcal{D}} 

\newcommand{\call}{\mathcal{L}}
\newcommand{\calu}{\mathcal{U}}
\newcommand{\calp}{\mathcal{P}}
\newcommand{\calq}{\mathcal{Q}}
\newcommand{\calr}{\mathcal{R}}
\newcommand{\calv}{\mathcal{V}}
\newcommand{\calt}{\mathcal{T}}
\newcommand{\calx}{\mathcal{X}}

\renewcommand{\commutator}[2]{\slr{#1, #2}}
\newcommand{\plr}[1]{{\left(#1\right)}}
\newcommand{\slr}[1]{\left[#1\right]}

\newcommand{\expect}[1]{\langle #1 \rangle}
\newcommand{\clr}[1]{\left\{#1\right\}}
\newcommand{\ceiling}[1]{{\left\lceil #1 \right\rceil}}

\begin{document}

\title{Universal first-passage time statistics for quantum diffusion}
\author{Guido Ladenburger}
\author{Finn Schmolke}
\author{Eric Lutz}
\affiliation{Institute for Theoretical Physics I, University of Stuttgart, D-70550 Stuttgart, Germany}

\begin{abstract}
First-passage phenomena play a fundamental role in classical stochastic processes. 
We here exactly solve a quantum first-passage time problem for quantum diffusion driven by measurement noise, a
generalization of classical Brownian motion. 
Such continuous monitoring may trap the measured quantum system in a decoherence-free subspace, a fraction of the available state space that is isolated
from the surroundings, and thus plays an important role in quantum information science. 
We analytically determine the first-passage time distribution, whose form neither depends on the system Hamiltonian nor on the measurement operator, and is therefore universal. 
These results provide a general framework to investigate the first-passage statistics of diffusive quantum trajectories.
\end{abstract}

\maketitle 

Determining the first time a random variable reaches a given  value is of central importance in the theory of stochastic processes. 
Assume the state space of a system can be divided into two (or more) sets, such that one of the sets (A) may be reached from the other one (B), but not vice versa. The time the system needs to transition from a state in B to a state in A defines the first-passage (or first-hitting) time \cite{wei64,kam97,gar97,red01,sun06,bra13,pol16}.
 It is  a stochastic quantity owing to the presence of random fluctuations. The corresponding first-passage time statistics provide key insights into the kinetics of a process. For that reason, their investigation  has found widespread application in physics, chemistry, biology and engineering, for instance, in the description of thresholded events, such as reaction kinetics, noise-activated escape, molecular rupture, and structural failure \cite{wei64,red01,kam97,gar97,sun06,bra13,pol16}. 
Since the early computations of the first-passage time distribution for free Brownian motion by Schr\"odinger \cite{sch15} and Smoluchowski \cite{smo15}, the framework of first-hitting phenomena for classical systems driven by thermal fluctuations is by now well established \cite{wei64,red01,kam97,gar97,red01,sun06,bra13,pol16}.

Motivated by potential applications in quantum network theory and quantum information science \cite{xia20}, first-passage studies have  been extended to monitored quantum systems \cite{kro06,kro08,dha15,dha15a,sin15,fri17,thi18,yin19,kes21,yin23,kul23,tor23,wan24,yin25,kor06,wil08,bol14,kew24}. A case in point is the analysis of the  first-hitting  time  of quantum random walks, defined as the time needed to first detect the quantum walker at a given (lattice) position \cite{kro06,kro08,dha15,dha15a,sin15,fri17,thi18,yin19,kes21,yin23,kul23}. Such first-detection time statistics have recently been examined on a quantum computer \cite{tor23,wan24,yin25}. On the other hand, quantum first-passage times have also been investigated along quantum trajectories  for specific examples of continuously monitored systems  \cite{kor06,wil08,bol14,kew24}. In both cases,  the actual observation of the system requires a   measurement, which unavoidably affects its state through quantum backaction, and leads to quantum fluctuations \cite{wis09,bar09,jac14,jor24}. As a result, classical and quantum first-passage processes are intrinsically different. It is fair to say that the theory of quantum {first-passage} is less developed than its classical counterpart. In particular, exact analytical expressions for first-passage time distributions as well as  universal  results, independent of specific Hamiltonians, are missing.

We here exactly solve a first-passage problem for the general quantum diffusion  induced by the continuous measurement of a system through homodyne detection \cite{wis09,bar09,jac14,jor24}. This stochastic process is a generalization of classical Brownian motion that takes place in Hilbert space rather than in physical configuration space, and fully retains quantum coherence and measurement backaction  \cite{wis09,bar09,jac14,jor24}. We concretely analyze the time it takes for the system to  randomly reach a decoherence-free subspace that is isolated from the surroundings \cite{dua97,zan97,lid98,lid03,blu08}. Since they undergo unitary,  coherence-preserving evolution,  such subspaces have played an important role in quantum information science \cite{dua97,zan97,lid98,lid03,blu08,kwi00,kie01,vio01,moh03,oll03,lan05,pre07,yam08,mon09,xu12,rei16}, with experimental realizations  presented in Refs.~\cite{kwi00,kie01,vio01,moh03,oll03,lan05,pre07,yam08,mon09,xu12,rei16}. We analytically evaluate the first-passage time distribution by mapping the quantum evolution onto a classical diffusion process with multiplicative noise, and explicitly compute its mean and variance. These results  neither depend on the system Hamiltonian nor on the measurement operator, and can hence be considered generic. We apply them to the examples of quantum nondemolition measurements \cite{gue07,bau11,ben14,bra80,bra96} and   of quantum synchronization \cite{buc22,sch22,sch24,tao25}.

\textit{Hilbert space structure  of quantum diffusion.} 
We  begin with the characterization of the Hilbert space structure of continuously monitored quantum systems, which corresponds to the sample space of the first-passage  problem.
Quantum diffusion, for a  system with Hamiltonian $H$ and state $\rho_W$, is often realized via homodyne detection, which allows a weak, ongoing observation of its state  without strong projection \cite{wis09,bar09,jac14,jor24}. The resulting time evolution is described by the stochastic master equation
\begin{align}\label{1}
    \dd{\rho}_W 
    =& -\I\commutator{H}{\rho_W} \dd{t} 
    +  \plr{L\rho_W L^\dagger-\frac{1}{2}\left\{L^\dagger L,\rho_W\right\}} \dd{t} \nonumber \\
    &+ \plr{{L}{\rho}_W + {\rho}_W L^\dagger - \expect{L + L^\dagger}{\rho}_W}\dd{W(t)},
\end{align} 
where $L$ is a general measurement operator, $\dd{W(t)}$ is a Wiener noise increment satisfying $\dd{W(t)^2} = \dd{t}$, and 
$\expect{\cdot} = \tr[\cdot {\rho}_W(t)]$ denotes the expectation value \cite{wis09,bar09,jac14,jor24}. To simplify the presentation, we consider a single measurement operator, but the analysis {is} easily extended to an arbitrary number of them (Supplemental Material). The first  observations of quantum trajectories were reported in superconducting qubits \cite{mur13,web14}. The density operator $\rho_W$ specifies a particular, noisy realization of the quantum process.
Taking the ensemble average over all trajectories, \cref{1} reduces to the usual Lindblad master equation,
$\dot{\rho} = -\I[H,\rho] + {D}[L]\rho$, where $\rho= \mathbb{E}[\rho_W]$ is the ensemble averaged density operator, and {${D}[L]\cdot = L\cdot L^\dagger - \left\{L^\dagger L,\cdot\right\}/2$} is the dissipator that accounts for the nonunitary dynamics generated by the coupling to the surroundings \cite{wis09,bar09,jac14,jor24}.

For systems obeying a Lindblad master equation, each Lindblad generator is related to a unique decomposition of the  Hilbert space into mutually orthogonal subspaces, $\mathcal{H} = \mathcal{D} \oplus \mathcal{R}$, where $\mathcal{D}$ is the subspace containing all the decaying states and $\mathcal{R}$ is the complement that contains all the states that do not decay completely \cite{bau08,bau12,car16}. 
The latter  correspond to the asymptotic states of the dynamics \cite{bau08,bau12}. 
{Such partition also holds at the level of single trajectories \cite{mou25,sch25}.}
In general, there is an additional finer Hilbert space structure such that the non-decaying space $\calr$ is itself composed of orthogonal subspaces \cite{bau08,bau12}. 
In particular, the presence of dynamical symmetries in the system-environment interaction \cite{dub23} leads to decoherence-free subspaces $\calq_\alpha$ for which the dissipator vanishes, $D[L]\rho_\calq =0$ for $\rho_\calq \in \calq_\alpha$ \cite{dua97,zan97,lid98,lid03,blu08}. 
These subspaces are composed of simultaneous eigenstates of both  Hamiltonian and Lindblad operators. In this case, the non-decaying subspace can be written as $\calr = \calq \oplus \calp$ with $\calq = \bigoplus_\alpha \calq_\alpha$ and $\calp = \calq^\perp$ is its orthogonal complement spanned by states that are not decoherence-free. 

A decoherence-free subspace remains decoherence-free along a single quantum trajectory, since according to \cref{1} $\dd{\rho_{W_\calq}} = -\I[H,\rho_{W_\calq}] \dd{t}$. 
Once a system is fully contained in a decoherence-free subspace (with probability one), it  cannot exit it anymore. The system is thus trapped in the subspace, which can accordingly be regarded  as a generalization of an absorbing state \cite{wei64,red01,kam97,gar97,red01,sun06,bra13,pol16}. The time needed to randomly end in a given decoherence-free subspace, driven by quantum (measurement) noise, therefore naturally defines a quantum first-passage time problem, extending the familiar classical concept to individual stochastic  trajectories in Hilbert space.

\begin{figure*}[t]
	\centering
	\begin{tikzpicture}
	\node (a) [label={[label distance=-.4 cm]145: \textbf{a)}}] at (0,0) {\includegraphics[width=0.3\textwidth]{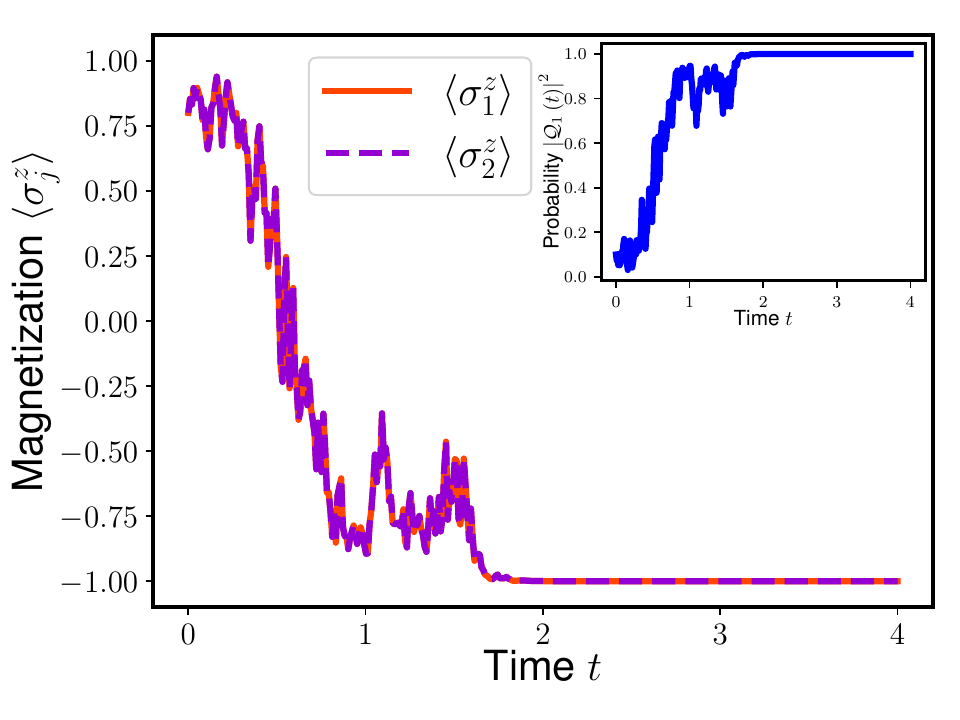}};	\node (a) [label={[label distance=-.4 cm]145: \textbf{b)}}] at (6,0) {\includegraphics[width=0.3\textwidth]{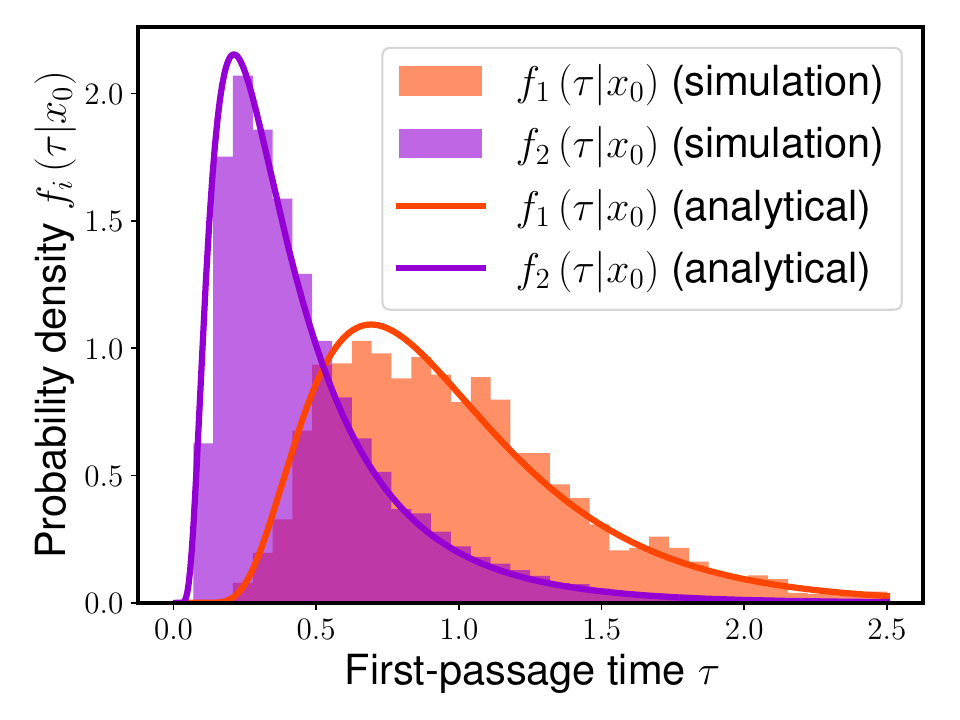}};
	\node (a) [label={[label distance=-.4 cm]145: \textbf{c)}}] at (12.0,0) {\includegraphics[width=0.3\textwidth]{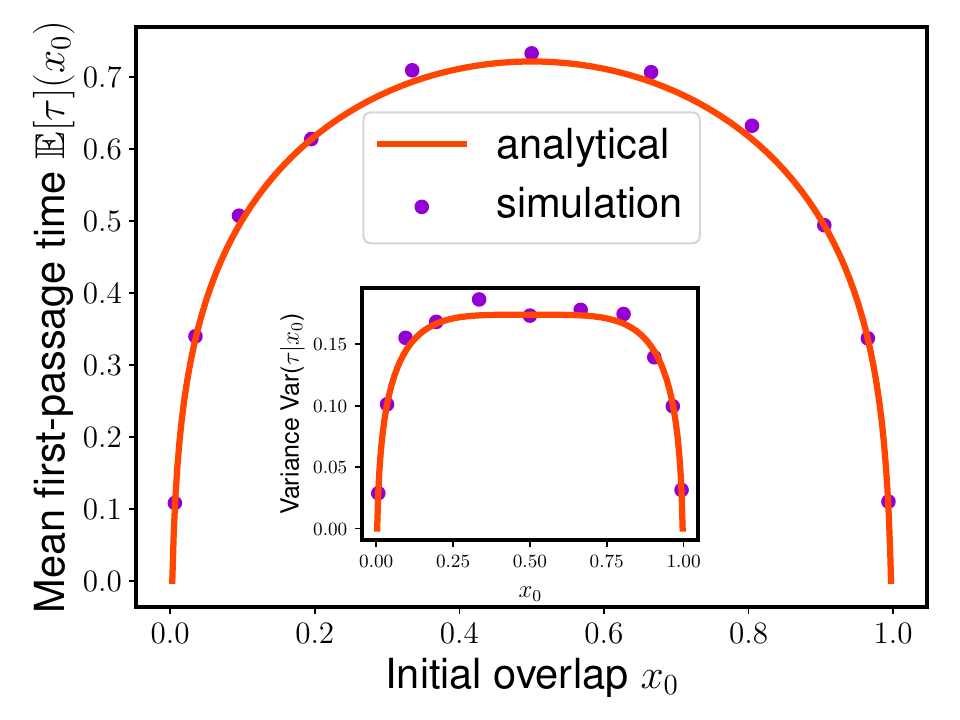}};
	\end{tikzpicture}
	\caption{Quantum nondemolition measurement of a two-qubit system. a) The magnetizations of the two spins along a single noisy trajectory are identical, even though only the first qubit is measured.  They stay constant after the system has reached the decoherence-free subspace  $\mathcal{Q}_1 = \text{span}(\ket{00})$. The inset shows the probability  to be in $\mathcal{Q}_1$. b) First-passage time distributions $ f_i(\tau|x_0)$, Eqs.~(7),  to be in  subspace $\mathcal{Q}_i$ $(i=1,2)$ with probability $1-\varepsilon$, for $\varepsilon = 0.003$ and initial overlap $x_0=0.1$ ($h_0=h_1=1$). Good agreement is obtained with the  simulations of the stochastic master equation (1) with $3\times 10^4$ trajectories. c) Mean and variance (inset) as a function of  the  initial probability $x_0$, Eqs.~(8)-(9), and simulated for $5\times 10^3$ trajectories.}\label{fig1}
\end{figure*}

\textit{Quantum first-passage time distribution}.  We next derive an analytic expression for the quantum first-passage time distribution of reaching a given  decoherence-free subspace. To that end, we introduce the probability to find the system in the decoherence-free subspace $\calq$, $|{\calq}(t)|^2 = \tr[\rho_W(t) P_{\calq}]$, where $P_\calq$ is the projector on that subspace, and examine its time evolution. This will allow us to reduce the quantum first-passage problem to a classical one \cite{wei64,red01,kam97,gar97,red01,sun06,bra13,pol16}. Using the stochastic master equation \eqref{1}, we obtain  (Supplemental Material)
\begin{align}\label{2}
\dd\plr{\abs{\calq(t)}^2} 
=& \sum_\alpha \abs{\calq_\alpha(t)}^2 \bigg([1 - \abs{\calq_1(t)}^2]2\m{Re}(c_\alpha) \nonumber \\
&- \tr[\rho_W(L_k+L_k^\dagger)P_\calp] \bigg) \dd W(t),
\end{align}
where $c_\alpha$ are the eigenvalues of the measurement operator, defined through {$P_{\calq_\alpha} L P_{\calq_\alpha} = c_\alpha P_{\calq_\alpha}$}, with $P_{\calq_\alpha}$ and $P_\calp$ the respective projectors on the  subspace $\calq_\alpha$ and the complement $\calp$. 
We note that the overlap with the complement vanishes, $\tr[\rho_W P_\calp] = 0$, when the system is in $\calq$.
We further assume that there are only two such subspaces, that is, $\calq = \calq_1 \oplus \calq_2$ (the case of multiple decoherence-free subspaces is addressed in the Supplemental Material).
The probability to find the system in  $\calq_1$ then follows as
\begin{equation}\label{3}
  \dd\plr{\abs{\calq_1(t)}^2} = 2\plr{c_1 - c_2} \abs{\calq_1(t)}^2
  \plr{1-\abs{\calq_1(t)}^2} \dd W.
\end{equation}
{\Cref{3} is a closed stochastic differential equation of the form $\dd x(t) = \sqrt{2 D(x)}\dd W(t)$, for the variable $x =\abs{\calq_1(t)}^2$, with a position-dependent diffusion coefficient $D(x) = 2\gamma^2 x^2 \plr{1-x}^2$.}
The quantity $\gamma = c_1-c_2$ can be viewed as a measure of the distinguishability of  the two decoherence-free subspaces; it plays the role of the effective strength of the diffusion coefficient. 
The above (classical) drift-free It\=o process with multiplicative noise can alternatively be described by a Fokker--Planck equation for the probability density $p(x,t)$ \cite{kam97,gar97}
\begin{equation}\label{4}
\partial_t p(x,t) 
  = \partial^2_x[D(x) p(x,t)] \!
  = \!2\gamma^2 \partial_x^2[x^2(1-x)^2 p(x,t)].\!
\end{equation}

Equation \eqref{4}  corresponds to a so-called degenerate diffusion process \cite{hui07,che20}, with a
  coefficient $D(x)$ that vanishes at the boundaries of the  diffusion interval, meaning that diffusion stops at these two points. These processes also occur in population genetics \cite{ewa04},  epidemiology \cite{cha14}, porous media clogging models \cite{sch17} and mathematical finance \cite{alm02}.  Their first-passage time properties depend on how fast the diffusion coefficient vanishes. Their boundary behavior can be  classified following Feller's criteria  \cite{fel54,kar81,eth86}: In the present case, the quadratic dependence close to $x=0 \text{ or }1$ leads to (natural) inaccessible boundaries \cite{che20} that cannot be reached in finite time (all {other} points of the interval $[\varepsilon, 1-\varepsilon]$, even arbitrary close to the boundaries ($\varepsilon \rightarrow 0$), are reached with probability one). The process hence asymptotically approaches, but never actually reaches the boundary in finite time. The mean first-passage time for the full interval is accordingly infinite. Divergent mean first-passage times are well-known in classical first-passage problems; this is, for example, the case of standard Brownian motion, where the boundaries at $\pm\infty$ are also  unattainable in finite time \cite{wei64,red01,kam97,gar97,red01,sun06,bra13,pol16}. However, they have profound consequences in quantum theory. As we will see below, they imply that projective measurements, as the limiting process of repeated weak measurements, as, for instance, experimentally implemented with quantum nondemolition measurements of photons in Ref.~\cite{gue07}, cannot, fundamentally (even for ideal measurements with unit efficiency \cite{wis09,bar09,jac14,jor24}), be implemented with probability one, but only with a finite fidelity that can be very close, but not equal, to one.

We proceed by solving \cref{4} with the help of the separation ansatz {$p(x,t) = \sum_n A_n F_n(x) \exp(-2\gamma^2 \lambda_n t)$.}
We impose the absorbing boundary conditions $p(\varepsilon, t) = p(1-\varepsilon, t) = 0,\ \varepsilon\in\plr{0, 1/2}$, and the initial condition $p(x,0) = \delta(x-x_0),\ x_0\in [\varepsilon, 1-\varepsilon]$ \cite{kar81}. 
We explicitly find with $n=1,2,\ldots$
\begin{widetext}
  \begin{equation}
    F_n(x) = \frac{\plr{-1}^\ceiling{\frac{n}{2}}}{\sqrt{\ln\frac{1-\varepsilon}{\varepsilon}}} \plr{x-x^2}^{-\frac{3}{2}} \sin(\frac{\pi}{2}n\slr{1+\plr{\ln\frac{\varepsilon}{1-\varepsilon}}^{-1}\ln\frac{1-x}{x}})\quad \text{and} \quad \lambda_n = \frac{1}{4}\left[1+\plr{\frac{\pi n}{\ln\frac{\varepsilon}{1-\varepsilon}}}^2\right].
  \end{equation}
\end{widetext}
Combining the initial condition and the orthonormality  of the {functions} $F_n(x)$ further leads to the coefficients $A_n = D(x_0) F_n(x_0)/2\gamma^2$ (Supplemental Material).
The  solution of the Fokker--Planck  equation \eqref{4} is therefore
\begin{equation}\label{6}
  p(x,t) = \frac{1}{2\gamma^2} D(x_0) \sum_n F_n(x_0)F_n(x) e^{-2\gamma^2 \lambda_n t}.
\end{equation}

\begin{figure*}[t]
	\centering
	\begin{tikzpicture}
	\node (a) [label={[label distance=-.4 cm]145: \textbf{a)}}] at (0,0) {\includegraphics[width=0.3\textwidth]{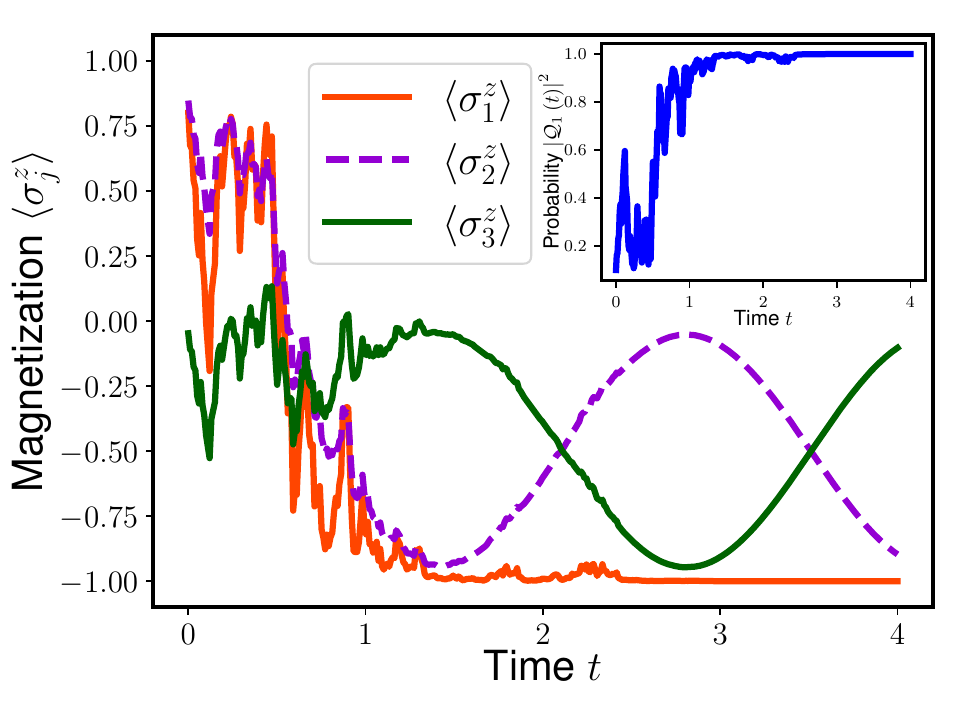}};	\node (a) [label={[label distance=-.4 cm]145: \textbf{b)}}] at (6,0) {\includegraphics[width=0.3\textwidth]{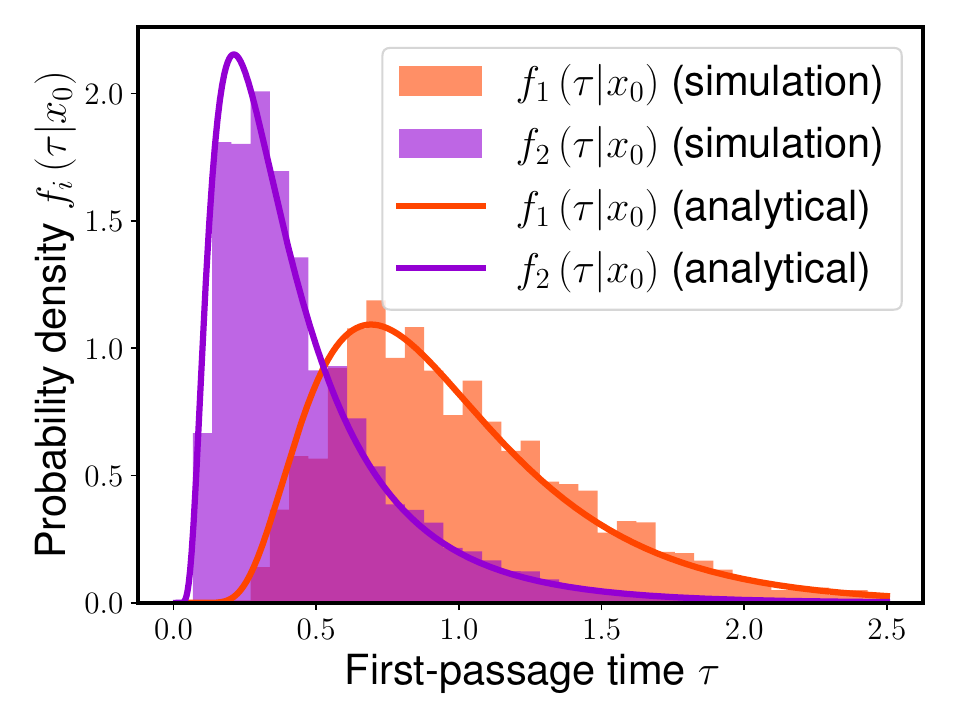}};
	\node (a) [label={[label distance=-.4 cm]145: \textbf{c)}}] at (12.0,0) {\includegraphics[width=0.3\textwidth]{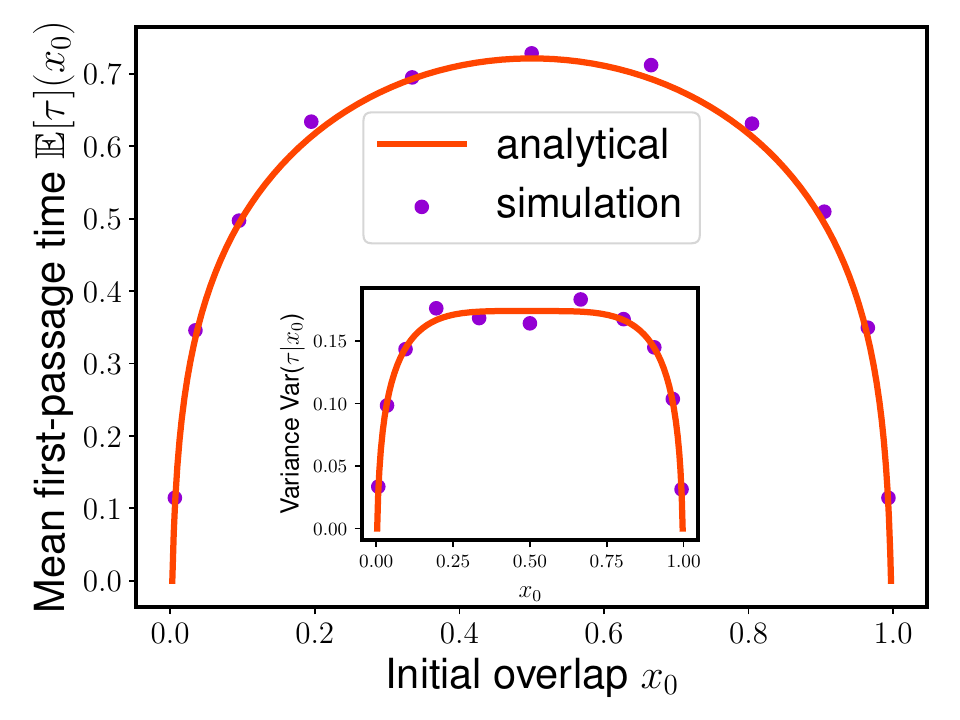}};
	\end{tikzpicture}
	\caption{Measurement-induced quantum synchronization for a five-qubit ring. a) The magnetizations of qubit 2 and qubit 3 transition from a noisy unsynchronized state to a noise-free  antisynchronized state after the system has reached the decoherence-free subspace $\mathcal{Q}_1$, when the first qubit is measured.   The inset shows the corresponding probability  to be in $\mathcal{Q}_1$. b) First-passage time distributions $ f_i(\tau|x_0)$, Eqs.~\eqref{7},  to be in  subspace $\mathcal{Q}_i$ $(i=1,2)$ with probability $1-\varepsilon$, for $\varepsilon = 0.003$ and initial overlap $x_0=0.1$ ($h_0=h_1=1$). Good agreement is obtained with the  simulations of the stochastic master equation \eqref{1} with $3\times 10^4$ trajectories. c) Mean and variance (inset) as a function of  the  overlap $x_0$, \cref{8,9}, and simulated for $5\times 10^3$ trajectories.}\label{fig2}
\end{figure*}

We may now solve the quantum first-passage time problem  using standard techniques \cite{wei64,red01,kam97,gar97,red01,sun06,bra13,pol16}. 
The first-passage time distribution $f_1(\tau|x_0)$ of reaching the decoherence-free space $\calq_1$ (with probability $1-\varepsilon$), starting at $x=x_0$, in other words, the probability to cross the upper boundary at $x=1-\varepsilon$ at time $\tau$, is $f_1(\tau|x_0)= -D(1-\varepsilon) \partial_{x^\prime}p(x^\prime,\tau)|_{x^\prime=1-\varepsilon}$ \cite{gar97}. Similarly, the first-passage time distribution $f_2(\tau|x_0)$ of reaching the decoherence-free space $\calq_2$ (with probability $1-\varepsilon$), starting at $x= x_0$, that is, the probability to cross the lower boundary at $x=\varepsilon$ at time $\tau$, is $f_2(\tau|x_0)= D(\varepsilon) \partial_{x^\prime}p(x^\prime,\tau)|_{x^\prime=\varepsilon}$ \cite{gar97}. Using the solution \eqref{6}, we concretely obtain
\begin{align}\label{7}
  f_1(\tau|x_0) &= -\frac{D(x_0) D(1-\varepsilon)}{2\gamma^2}  \sum_{n=1}^{\infty} F_{n}(x_0) F_{n}^\prime(1-\varepsilon) e^{-2\gamma^2 \lambda_{n}\tau},\notag\\
  f_2(\tau|x_0) &= \frac{D(x_0) D(\varepsilon)}{2\gamma^2}  \sum_{n=1}^{\infty} F_{n}(x_0) F_{n}^\prime(\varepsilon) e^{-2\gamma^2 \lambda_{n}\tau}.
\end{align}
Expressions \eqref{7} fully solve the formulated  first-passage time problem for quantum diffusion. It is important to stress that the above analytical distributions are exact; exact analytical first-passage time distributions are rare, even in classical physics \cite{wei64,red01,kam97,gar97,red01,sun06,bra13,pol16}. The overall form of these two distributions neither depends on the system Hamiltonian $H$ nor on the measurement operator $L$, and is hence universal.
{It is, in particular, independent of the Hilbert space dimension.}
Information about the specific system at hand is only encoded in the parameter $\gamma$.
 
Since the It\=o  process \eqref{3} is drift-free, the mean first-passage time $\mathbb{E}[\tau](x_0)$ can be  easily evaluated from  the equation $ D(x_0) \partial^2_{x_0}\mathbb{E}[\tau](x_0) = -1$ \cite{kam97,gar97}. Imposing the boundary conditions $\mathbb{E}[\tau](\varepsilon) = \mathbb{E}[\tau](1-\varepsilon) = 0$, we find 
\begin{equation}\label{8}
    \mathbb{E}[\tau] \!=\! \eta(x_0) - \eta(\varepsilon) \text{ with }  \eta(x) \!= \!\frac{1}{\gamma^2}\!\left(\!x-\frac{1}{2}\!\right)\!\ln(\!\frac{1}{x}-1\!)\!\!\!
\end{equation}
Equation \eqref{8} has a surprisingly simple, universal form. The mean first-passage time only depends  on the initial overlap $x_0$ with the decoherence-free subspace and the chosen threshold value $\varepsilon$.
Details about the continuously monitored quantum system again only enter via the constant $\gamma$.
Stronger measurement and higher distinguishability between the subspaces both reduce the mean first-passage time by increasing $\gamma$.
Although \cref{3} has the form of classical stochastic processes,  it  fundamentally differs from the classical theory  \cite{wei64,red01,kam97,gar97,red01,sun06,bra13,pol16}.
The variable $x = |\calq_1(t)|^2$ indeed corresponds to the quantum probability of finding the system in a particular corner of the Hilbert space, and is thus inherently a nonlocal quantity that accommodates linear superpositions of states.  Expression \eqref{8} additionally reveals that the mean first-passage time diverges logarithmically when the boundaries of the diffusion interval are approached.

The variance  $ \text{Var}(\tau\vert x_0) =\mathbb{E}[\tau^2](x_0) - \mathbb{E}[\tau]^2(x_0)$ can be similarly determined by integrating  $  D(x_0)\partial^2_x\mathbb{E}[\tau^2](x_0) = -2\mathbb{E}[\tau]$ with $\text{Var}({\tau\vert \varepsilon}) = \text{Var}({\tau\vert 1-\varepsilon}) = 0$ \cite{kam97,gar97}.
We have
\begin{align}
  \text{Var}({\tau\vert x_0}) 
  =& \frac{1}{\gamma^2}\mathbb{E}[\tau] + \eta^2({\varepsilon}) - \eta^2({x_0})\nonumber\\
  &- \frac{1}{4\gamma^4}\left[\ln(\frac{1}{x_0}-1)^2 - \ln(\frac{1}{\varepsilon}-1)^2\right],
\label{9}
\end{align}
with the same function $\eta(x)$ as in \cref{8}.

\textit{Illustration 1: Quantum nondemolition measurement}.  We now illustrate the above general results  by analyzing two   examples. We first consider an elementary model of repeated quantum nondemolition measurements \cite{bau11}  on a two-qubit system with Hamiltonian $ {H} = h_0 \plr{{\sigma}_1^z + {\sigma}_2^z} + h_1 \plr{{\sigma}_1^+{\sigma}_2^- + {\sigma}_1^-{\sigma}_2^+}$. A sufficient criterion for quantum nondemolition measurements is that the Hamiltonian commutes with the measurement operator \cite{bra80,bra96}. We concretely continuously measure the first qubit with the measurement operator $L= \sigma^z_1$. This system has two decoherence-free subspaces  that are spanned by $\ket{00}$ (subspace $\mathcal{Q}_1$) and $\ket{11}$ (subspace $\mathcal{Q}_2$), where $\ket{0}$ and $\ket{1}$ denote the respective ground and excited states of each qubits. The two eigenvalues of $L$ are respectively $c_1 = -1$ and $c_2 = 1$. We moreover choose the  initial state
 $\ket{\psi_0} = \sqrt{x_0}\ket{00} + \sqrt{1-x_0}\ket{11}$, with overlap $x_0$ ($1-x_0$) with the first (second) subspace.
  
 \Cref{fig1}a shows the evolution of the magnetizations $\langle \sigma^z_j\rangle $ of the two qubits as the system reaches the subspace $\mathcal{Q}_1$ for $\varepsilon = 0.003$; the inset displays the  probability $x = | {\mathcal Q}_1(t)|^2$. Remarkably, the two qubits behave in an identical  way, even though only the first one is measured. \Cref{fig1}b further presents the first-passage distributions $ f_1(\tau|x_0)$ and $ f_2(\tau|x_0)$, Eqs.~\eqref{7}, for $x_0=0.1$, while \cref{fig1}c shows the associated mean and variance (inset), Eqs.~\eqref{8}-\eqref{9}, as a function of $x_0$.  In all cases, we observe excellent agreement between the general analytical predictions and the simulations of the quantum trajectories according to \cref{1} using Python's {QuTiP package \cite{qutip1,qutip2,qutip3}}. 
 
Repeated quantum nondemolition measurements asymptotically converge towards projective {measurements \cite{gue07,bau11,ben14}.} They thus provide a microscopic mechanism for strong measurements. The mean first-passage time corresponds in this instance to the  time to project  the  two qubits   in  $\ket{00}$ or $\ket{11}$ with precision $\varepsilon$, that is, to the measurement duration. The associated temporal cost can be quantified via a general time-fidelity trade-off relation given by \cref{8}: the (average) time needed to achieve a given fidelity $F(\rho_W,\dyad{q}) = 1-\varepsilon$ with a state $\ket{q}$ indeed satisfies the general inequality
 $ \mathbb{E}[\tau] \ge \eta(x_0) + [\left(F-1/2\right)/\gamma^2] \ln[F/(F-1)]$; equality is only reached for unit detector efficiency (Supplemental Material).
The results of the (von Neumann) projective measurement \cite{aul09} are hence never reached with probability one in finite time.
 
 \textit{Illustration 2: Measurement-induced quantum synchronization.}  The second example consists of  a ring of five  coupled qubits with Hamiltonian
            ${H} = h_0\sum_{i=1}^5 \sigma_i^z + h_1\sum_{i=1}^5\plr{{\sigma}_i^+{\sigma}_{i+1}^- + {\sigma}_i^-{\sigma}_{i+1}^+}$, where the first qubit is measured with operator ${L} = {\sigma}_1^z$. This system bears some resemblance with the previous example, however,  the larger Hilbert space leads to new physics. It possesses two {distinct} decoherence-free subspaces: the subspace $\mathcal{Q}_1 = \text{span}(\ket{q_{11}},\ket{q_{12}},\ket{q_{13}})$ spanned by the three states $\ket{q_{11}}=\sqrt{2/5}\sum_{n=0}^{4}\sin(2\pi n/5)\ket{1}_n$, $\ket{q_{12}}=\sqrt{2/5}\sum_{n=0}^{4}\sin(4\pi n/5)\ket{1}_n$ and $\ket{q_{13}} = \ket{0}^{\otimes 5}$ associated with the ${L}$-eigenvalue $c_1=-1$, and the subspace $\mathcal{Q}_2 = \text{span}({\ket{q_{21}},\ket{q_{22}},\ket{q_{23}}})$, spanned by the three states $\ket{q_{21}}=\sqrt{2/5}\sum_{n=0}^{4}\sin(2\pi n/5)\ket{0}_n$, $\ket{q_{22}}=\sqrt{2/5}\sum_{n=0}^{4}\sin(4\pi n/5)\ket{0}_n$ and $\ket{q_{23}} = \ket{1}^{\otimes 5}$ corresponding to the ${L}$-eigenvalue $c_2=1$.
            
            \Cref{fig2}a exhibits the magnetizations of the first three qubits as a function of time, for the initial state $\ket{\psi_0} = [\sqrt{x_0}(\ket{q_{11}} + \ket{q_{12}}) + \sqrt{1-x_0}(\ket{q_{21}} + \ket{q_{22}})]/\sqrt{2}$ and $\varepsilon = 0.003$. A spontaneous transition between a noisy, unsynchronized phase and a deterministic, synchronized phase where the oscillations of qubits 2 and  3 are perfectly (anti)synchronized occurs, when the system gets trapped in  the subspace $\mathcal{Q}_1$ (inset). As before, the simulated  first-passage distribution based on \cref{1}, as well as  mean and variance,   match the analytical  predictions, Eqs.~\eqref{7}-\eqref{9}, (\cref{fig2}bc). Quantum synchronization is a generalization of classical synchronization  \cite{pik03,bal09}, where  expectations of  quantum observables oscillate in (anti)phase \cite{buc22,sch22,sch24,tao25}. The mean first-passage time  hence gives  here the synchronization time, that is, the average time before synchronization sets in with fidelity $1-\varepsilon$. 

\textit{Conclusions}. We have developed  a general framework to investigate the  first-passage statistics of diffusive quantum trajectories towards attracting subspaces, such as decoherence-free subspaces and, by extension, their special cases, {dark states \cite{scu10}}. By mapping the time evolution of the subspace probability to a classical Fokker--Planck equation with multiplicative noise, we have derived exact analytical expressions for the quantum first-passage time distribution and its lowest cumulants.  An important physical insight  is that  decoherence-free subspaces can only be reached with a probability $1-\varepsilon$, with $\varepsilon$ arbitrarily small, but nonzero, in finite time. The mean first-passage time diverges logarithmically as $\varepsilon$ tends to zero, leading to a time-fidelity trade-off. Our findings  are able to accurately capture the first-hitting times of {different} quantum stochastic processes, from quantum nondemolition measurements to quantum synchronization, and should hence be useful to describe the general kinetics of continuously monitored many-qubit systems \cite{maj07,roc14}.

\begin{acknowledgements} \textit{Acknowledgements.} We acknowledge financial support from the German Research Foundation (FOR 2724).

\end{acknowledgements}

\clearpage
\widetext
\begin{center}
\textbf{\large Supplemental Material: Universal first-passage time statistics for quantum diffusion}
\end{center}
\setcounter{equation}{0}
\setcounter{figure}{0}
\setcounter{table}{0} 
\setcounter{page}{1}
\makeatletter
\renewcommand{\theequation}{S\arabic{equation}}
\renewcommand{\thefigure}{S\arabic{figure}}
\renewcommand{\bibnumfmt}[1]{[S#1]}
\renewcommand{\citenumfont}[1]{S#1}


\renewcommand{\figurename}{Supplementary Figure} 
\renewcommand{\theequation}{S\arabic{equation}}
\renewcommand{\thefigure}{S\arabic{figure}}
\renewcommand{\bibnumfmt}[1]{[S#1]}
\renewcommand{\citenumfont}[1]{S#1}
\setcounter{secnumdepth}{3}
\makeatletter
\def\@seccntformat#1{\csname the#1\endcsname.\quad}
\makeatother
\renewcommand{\thesection}{\Roman{section}}
\thispagestyle{empty}

The Supplemental Material contains details about  (I)  the structure of the Hilbert space of an open quantum system induced by a general Lindblad generator, (II) the derivation of the stochastic differential equation for the overlap with an orthogonal subspace, (III) the explicit solution of the classical Fokker--Planck equation for transitions between decoherence-free subspaces, (IV)  the analytic derivation of the  first-passage time distribution, and (V) the computation of the corresponding mean first-passage time and  variance.

\section{Hilbert space structure}
\label{sec:structure}
Our starting point is an otherwise closed quantum system subject to indirect continuous monitoring, with an evolution described by the \Ito stochastic master equation \cite{wis09s,bar09s,jac14s,jor24s}
\begin{align}
  \dd{\rho}_W 
  = -\I[H,\rho_W] \dd{t}
  + \sum_k \left(L_k \rho_W L^\dag_k - \frac{1}{2}\left\{L^\dag_kL_k,\rho_W\right\}\right) \dd{t}
  + \plr{L_k\rho_W + {\rho}_W L_k^\dagger - \expect{L_k + L_k^\dagger} \rho_W}\dd{W_k}.
  \label{eq:diffusive-trajectory}
\end{align}
The operators $L_k$ mediate the effective action of the different indirect measurement processes with independent Wiener increments $\dd{W_k}(t)$, satisfying $\mathbb{E}[\dd{W_k(t)}] = 0$ and $\mathbb{E}[\dd{W_j(t)}\dd{W_k(t^\prime)}] = \delta_{jk}\delta(t-t^\prime)\dd{t}$.
Taking the average over the ensemble of trajectories results in the mean evolution of the density operator $\rho = \mathbb{E}[\rho_W]$ which is governed by the Lindblad master equation
\begin{align}
  \dot{\rho} = \call(\rho) 
  = -\I[H,\rho]
  + \sum_k \left(L_k \rho L^\dag_k - \frac{1}{2}\left\{L^\dag_kL_k,\rho\right\}\right).
  \label{eq:Lindblad-equation}
\end{align}
Denote by
    $\calt^t = \exp(\call t)$  
the quantum dynamical semigroup associated with the Lindblad generator $\call$.
In general, any Lindblad generator imposes a structuring upon Hilbert space into mutually orthogonal subspaces \cite{bau08s,bau12s,car16s}.
While the Lindblad equation evolves orthogonal subspaces independently, quantum trajectories are susceptible to the Hilbert space structure and become asymptotically confined to only a fraction of the available state space \cite{ben25s,sch25s}.

There are two levels to this decomposition.
The entire Hilbert space can always be uniquely partitioned into two macroscopic subspaces according to
\begin{equation}
  \mathcal{H} = \mathcal{D} \oplus \calr
\end{equation}
where $\cald$ is the decaying subspace; it is the largest subspace that contains only decaying states
\begin{equation}
  \mathcal{D} = \clr{\ket{\psi}\in\mathcal{H} \left\vert \Braket{\psi \vert \mathcal{T}^t(\rho)\vert \psi} \xrightarrow{t\to\infty} 0 \ \forall\rho\right.},
\end{equation}
and gets completely emptied in the evolution.
Since any initial state eventually loses support on $\cald$, the presence of decay ($\cald \neq \emptyset$) is directly linked to rank-decreasing evolution.
The complement $\calr$ is the largest subspace that contains the support of all asymptotic states 
\begin{equation}
    \calr = \clr{\ket{\psi}\in\mathcal{H} \left\vert \Braket{\psi \vert \mathcal{T}^t(\rho)\vert \psi} \xrightarrow{t\to\infty} \text{const.} >0 \ \forall\rho\right.}.
\end{equation}
It may have inflow (if $\cald \neq \emptyset$) but no outflow of probability.
The asymptotic subspace $\calr$ has itself a microscopic substructure, it can be decomposed into its irreducible components \cite{bau12s} (Theorem 7)
\begin{equation}
  \calr 
  = \bigoplus_{k=1}^K \calu_k \oplus \bigoplus_{l=1}^M \calx_l, \quad 
  \calx_l = \bigoplus_{\nu=1}^{m(l)} \calv_{l,\nu} \simeq \mathbb{C}^{m(l)} \otimes \calv_l, \quad \calv_l \simeq \calv_{l,\nu},\ \forall \nu.
  \label{eq:R-decomposition-general}
\end{equation}
Each element $\mathcal{U}_k$ and $\calv_{l,\nu}$ is atomic in the sense that it cannot be further decomposed into smaller sets of orthogonal subspaces; they are the minimal orthogonal subspaces and correspond to the range of exactly one extremal stationary state of the Lindblad equation \eqref{eq:Lindblad-equation}.
The subspaces $\calu_k$ constitute the unique part of the decomposition \eqref{eq:R-decomposition-general}.
On the other hand, the subspace $\calx_l$ collects all minimal subspaces $\calv_{l,\nu}$ that have unitarily equivalent counterparts with $\calv_l \simeq \calv_{l,\nu},\ \forall \nu$ ($A \simeq B$ denotes the existence of an isomorphism between $A$ and $B$).
These are the degenerate subspaces, where $m(l)$ denotes the degree of degeneracy.

We are here particularly interested in systems that posses at least two different decoherence-free subspaces.
A decoherence-free subspace $\calq$ is a common invariant subspace of both the Hamiltonian $H$ and all jump operators $L_k$, on which
\begin{align}
    [(H)_\calq,(L_k)_\calq] = 0 \ \forall k, \quad [(L_j)_\calq,(L_k)_\calq] = 0 \ \forall j,k,
\end{align}
where $(\cdot)_\calq$ denotes the restriction to the subspace $\calq$.
The Hamiltonian restricted to $\calq$ is a strong symmetry \cite{buc12s} and consequently, $(H)_\calq$ and $(L_k)_\calq$ are simultaneously diagonal.
A macroscopic decoherence-free subspace may be composed of smaller ones $\calq = \bigoplus_\alpha \calq_\alpha$.
The different $\calq_\alpha$ can be identified as the common degenerate eigenspaces of the restricted jump operators $(L_k)_\calq$.
The decoherence-free states spanning the orthogonal subspaces $\calq_\alpha$ thus satisfy the condition
\begin{align}
  L_k \ket{q} = c_{k,\alpha} \ket{q}, \quad \forall \ket{q} \in \calq_\alpha,
  \label{eq:dfs-condition}
\end{align} 
with $c_{\alpha,k} \in \mathbb{C}$.
With respect to the decomposition \cref{eq:R-decomposition-general}, we can make the following identification
\begin{align}
  \calq_\alpha = 
  \begin{dcases}
    \calu_k, \ \m{dim}(\calq_\alpha) = 1,\\
    \calx_l, \ m(l) = \m{dim}(\calq_\alpha) > 1
  \end{dcases}.
\end{align}
In the following, we detail how trajectories react to the structure of the state space.

\section{Probabilities along quantum trajectories}
\label{sec:probabilities-along-quantum-trajectories}
Equipped with the structure of the state space, we next derive the evolution equation for the overlap of a quantum trajectory with an orthogonal subspace.
On the ensemble level, asymptotic dynamics takes place exclusively on the asymptotic subspace $\calr$.
It has been shown that also individual quantum trajectories eventually must converge to $\calr$ \cite{ben17,sch25s}.
It  is therefore sufficient to consider only times where the stochastic trajectory $\rho_W$ has already reached full support on $\calr$.

If a decoherence-free subspace $\calq$ exists, the asymptotic subspace may always be bipartioned according to \cite{bau08s,bau12s}
\begin{equation}
  \calr = \calq \oplus \mathcal{P},
\end{equation}
where $\mathcal{P} = \calq^\perp$ is the orthogonal complement, i.e. $\mathcal{P}$ is spanned by states that are not decoherence-free. 
From \cref{eq:diffusive-trajectory}, we obtain the evolution of the expectation value of an observable $A$,
\begin{equation}
  \dd\expect{A} = \tr(\dd\rho(t) A).
  \label{eq:evolution-expectation-value}
\end{equation}
In particular, we may choose $A$ to be the projector onto the subspace $\calq$, i.e. $A = P_\calq$, so that \cref{eq:evolution-expectation-value} describes the evolution of the probability to find the system in $\calq$ at time $t$, and we write for the stochastic differential
\begin{equation}
  \dd \big(|\calq(t)|^2\big) = \tr(\dd\rho(t)P_\calq).
  \label{eq:probability-differential}
\end{equation}
In general, the evolution of the overlap follows the stochastic differential equation \cite{sch25s}
\begin{align}
  \dd{\big(|\calq(t)|^2 \big)}&= \dd{(\tr[\rho_W(t)P_\calq])}
= \sum_k \bigg(|\calp(t)|^2\tr[\rho_W(t)(L_k+L_k^\dagger)P_\calq]
-|\calq(t)|^2 \tr[\rho_W(t)(L_k+L_k^\dagger)P_\calp]\bigg) \dd{W_k},
\label{eq:full-differential}
\end{align}
where $P_\calp$ is the projector onto the subspace $\calp$.
Assuming that the subspaces are identifyable by individual quantum trajectories, every realization will eventually converge to either the subspace $\calq$ with probability $\lim_{t\to\infty}\tr[\rho(t)P_\calq]$ or to the subspace $\calp$ with probability $\lim_{t\to\infty}\tr[\rho(t)P_\calp]$ \cite{ben25s,sch25s}.

We are here concerned only with the subensemble that undergoes transitions into the decoherence-free subspace $\calq$.
In the presence of multiple decoherence-free subspaces one may always perform an arbitrary bipartition of the whole collection $\calq = \calq_1 \oplus \calq_2$, where now both $\calq_j = \bigoplus_{\alpha=1}^{M_j} \calq_{j,\alpha}$, $j = 1,2$ contain in principle an arbitrary number of $M_j$ decoherence-free subspaces.
This problem reduces to a first-passage time problem between the two macroscopic decoherence-free subspaces $\calq_1$ and $\calq_2$ and using the property \eqref{eq:dfs-condition}, the differential  for the probability to find the system in $\calq_1$ becomes
\begin{align}
  \dd \big(|\calq_1(t)|^2\big) 
  = 2\sum_k \bigg(|\calq_2(t)|^2 \sum_\alpha \m{Re}(c_{k,\alpha})|\calq_{1,\alpha}(t)|^2
  - |\calq_1(t)|^2 \sum_\beta \m{Re}(c_{k,\beta})|\calq_{2,\beta}(t)|^2\bigg)\dd{W_k}.
\label{eq:dfs}
\end{align}
If there are only two different subspaces ($M_j = 1$), \cref{eq:dfs} can be reexpressed as
\begin{align}
  \dd \big(|\calq_1(t)|^2\big) 
  = 2\gamma|\calq_1(t)|^2\plr{1-\abs{\calq_1(t)}^2}\dd{W},
  \label{eq:dfs-closed}
\end{align}
where $\dd{W} = \sum_k \m{Re}(c_{k,1}-c_{k,2}) \dd{W_k}/\gamma$ is a standard Wiener process and $\gamma = \sqrt{\sum_k \m{Re}(c_{k,1}-c_{k,2})^2}$.
\Cref{eq:dfs-closed} reduces to Eq.~(2) of the main text if there is only one jump operator $L$.

Generalizing the above considerations to account for an imperfect detector efficiency is now immediate.
Suppose that only a fraction $\zeta \in [0,1]$ of the measurement results are captured by the detector.
The \Ito stochastic differential equation for a single measurement channel with operator $L$ becomes \cite{wis09s,jac14s}
\begin{align}
  \dd{\rho_W} = -\I [H,\rho_W]\dd{t} + \left(L\rho_W L^\dagger - \frac{1}{2}\left\{L^\dagger L,\rho_W\right\}\right) \dd{t} + \sqrt{\zeta}\left(L\rho_W + \rho_W L^\dag - \langle L+L^\dagger\rangle \rho_W\right) \dd{W}.
\end{align}
Going through the same steps as before directly yields 
\begin{align}
  \dd \big(|\calq_1(t)|^2\big) 
  = 2\tilde \gamma|\calq_1(t)|^2\plr{1-\abs{\calq_1(t)}^2}\dd{W},
\end{align}
with the modified parameter $\tilde \gamma = \sqrt{\zeta}\gamma$.
A nonideal detector thus effectively reduces the noise strength and increases the mean first-passage time (cf. \cref{eq:mean-fpt-reduced}).

\section{Solution of the Fokker--Planck equation}
In this section, we solve the Fokker--Planck equation for the probability distribution $p(|\calq_1|^2,t)$ subject to absorbing boundary conditions.
The first-passage time distribution then follows from $p(|\calq_1|^2,t)$  as detailed in \cref{sec:fpt}.

\Cref{eq:dfs-closed} is the stochastic differential equation of a drift-free It\=o diffusion process \cite{kam97s,gar97s}.
Introducing the shorthand notation $x = \abs{\calq_1(t)}^2$ yields the more familiar form
\begin{equation}
  \dd x = \sqrt{2 D(x)}\dd W,
  \label{eq:dx}
\end{equation}
where
\begin{equation}
  D(x) = 2\gamma^2 x^2 \plr{1-x}^2,
  \label{eq:diffusion-coefficient}
\end{equation}
is the state-dependent diffusion coefficient.
The dynamics of $x$ generated by \cref{eq:dx} has two fixed points at the boundaries, $x = 0,1$.
\Cref{eq:dx} can be equivalently described by the Fokker--Planck equation for the probability density $p(x,t)$ \cite{kam97s,gar97s},
\begin{equation}
  \pdv{p(x,t)}{t} = \pdv[2]{x} \plr{D(x) p(x,t)}
  = 2\gamma^2 \pdv[2]{x}\plr{x^2\plr{1-x}^2 p(x,t)}.
  \label{eq:fokker-planck-equation}
\end{equation}
This Fokker--Planck equation is linear and can thus be solved via spectral decomposition
using  the separation ansatz
\begin{equation}
  p(x,t) = \sum_n A_n F_n(x) \exp(-2\gamma^2 \lambda_n t),
  \label{eq:ansatz}
\end{equation}
with yet to be determined eigenfunctions $F_n(x)$ and  eigenvalues $\lambda_n$. The corresponding eigenvalue equation reads
\begin{equation}
  \pdv[2]{x} \plr{\frac{1}{2\gamma^2} D(x) F_n(x)} = -\lambda_n F_n(x).
  \label{eq:eigen-equation}
\end{equation}
The singular scaling of the diffusion coefficient at the boundaries  prevents trajectories from reaching either fixed point in finite time \cite{kar81s,eth86s}.
Any interval $[\varepsilon,1-\varepsilon]$ with $\varepsilon \in (0,1/2)$ is however accessible (with probability one) and we therefore impose the absorbing boundary conditions \cite{kar81s,eth86s}
\begin{equation}
  p(\varepsilon,t) = p(1-\varepsilon,t) = 0,\quad \varepsilon \in (0,1/2).
  \label{eq:absorbing}
\end{equation}
We choose the initial state to have a given initial overlap $x_0 = |\calq_1(0)|^2$, which directly translates to the initial condition
\begin{equation}
  p(x,0) = \delta(x-x_0), \quad x_0 \in [\varepsilon,1-\varepsilon].
  \label{eq:initial}
\end{equation}
The more general situation where the initial overlap has a distribution $p(x_0)$ can be straightforwardly accounted for \cite{gar97s}.
Such a situation may for instance arise if the collection of decoherence-free subspaces $\calq$ is reached following a cascade of localization transitions from an initially larger subspace where $x_0$ effectively becomes a random variable.

With the above boundary conditions, \cref{eq:eigen-equation} can be solved by the eigenfunctions
\begin{equation}
  F_n(x) = \frac{(-1)^\ceiling{\frac{n}{2}}}{\sqrt{\ln\frac{1-\varepsilon}{\varepsilon}}} (x-x^2)^{-\frac{3}{2}} \sin(\frac{\pi}{2}n\slr{1+\plr{\ln\frac{\varepsilon}{1-\varepsilon}}^{-1}\ln\frac{1-x}{x}}), \quad n=1,2,\ldots
\end{equation}
Their corresponding eigenvalues are given by
\begin{equation}
  \lambda_n = \frac{1}{4}\plr{1+\plr{\frac{\pi n}{\ln\frac{\varepsilon}{1-\varepsilon}}}^2}, \quad n=1,2,\ldots
\end{equation}
The eigenfunctions $F_n(x)$ form a complete basis on the interval $x \in [\varepsilon,1-\varepsilon]$ and satisfy the orthonormality condition
\begin{equation}
  \delta_{mn} = \int_\varepsilon^{1-\varepsilon}\dd{x} \frac{1}{2\gamma^2} D(x) F_m(x) F_n(x),
\end{equation}
with integration measure $\dd{x}D(x)/(2\gamma^2)$ \cite{art18s}.
Consequently, the expansion coefficients $A_n$ can be determined by applying the scalar product to \cref{eq:ansatz} at time $t=0$, to obtain
\begin{equation}
  A_n = \int_{\varepsilon}^{1-\varepsilon} \dd{x} \frac{1}{2\gamma^2} D(x) p(x,0) F_n(x).
\end{equation}
For an initially delta distributed overlap $p(x,0) = \delta(x-x_0)$ the coefficients reduce to 
\begin{align}
  A_n = \frac{1}{2\gamma^2} D(x_0) F_n(x_0).
\end{align}
The full solution of the Fokker--Planck equation \eqref{eq:fokker-planck-equation} subject to the absorbing boundary conditions \eqref{eq:absorbing} and the initial condition \eqref{eq:initial} is accordingly given by
\begin{equation}
  p(x,t) = \frac{1}{2\gamma^2} D(x_0) \sum_n F_n(x_0)F_n(x) e^{-2\gamma^2 \lambda_n t}.
  \label{eq:fpe-solution}
\end{equation}

\section{First-passage time distribution}
\label{sec:fpt}
\begin{figure}[t!]
        \centering       
        \begin{tikzpicture}
            \node (a) [label={[label distance=-.35 cm]140:\textbf{a)}}]  at (-12,0.) {\includegraphics[width=0.40\textwidth]{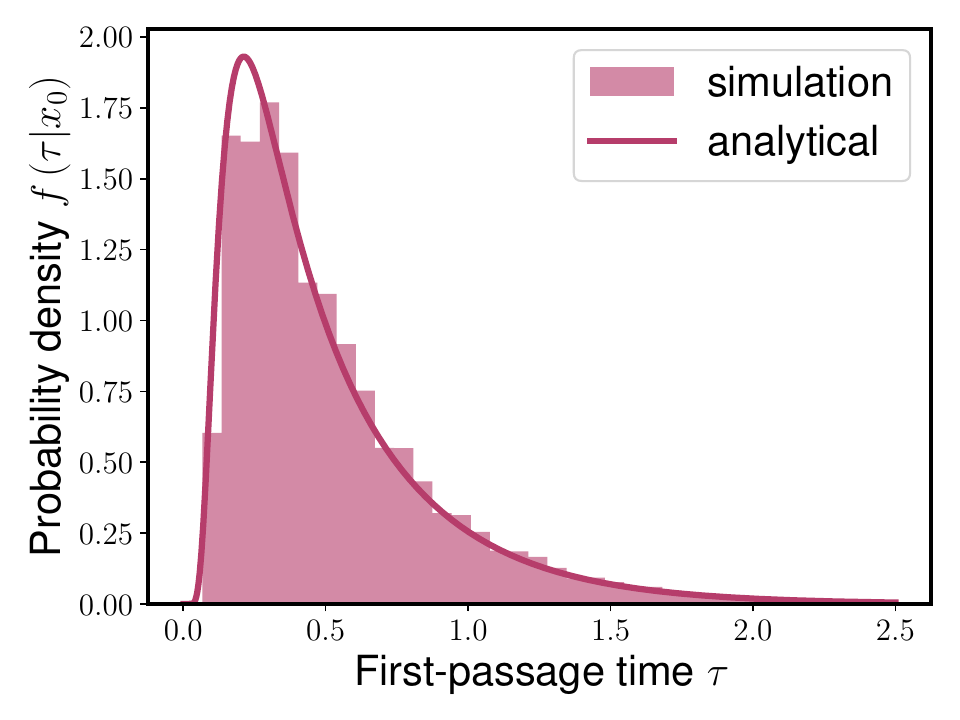}};
            \node (a) [label={[label distance=-.35 cm]140:\textbf{b)}}]  at (-4,0) {\includegraphics[width=0.40 \textwidth]{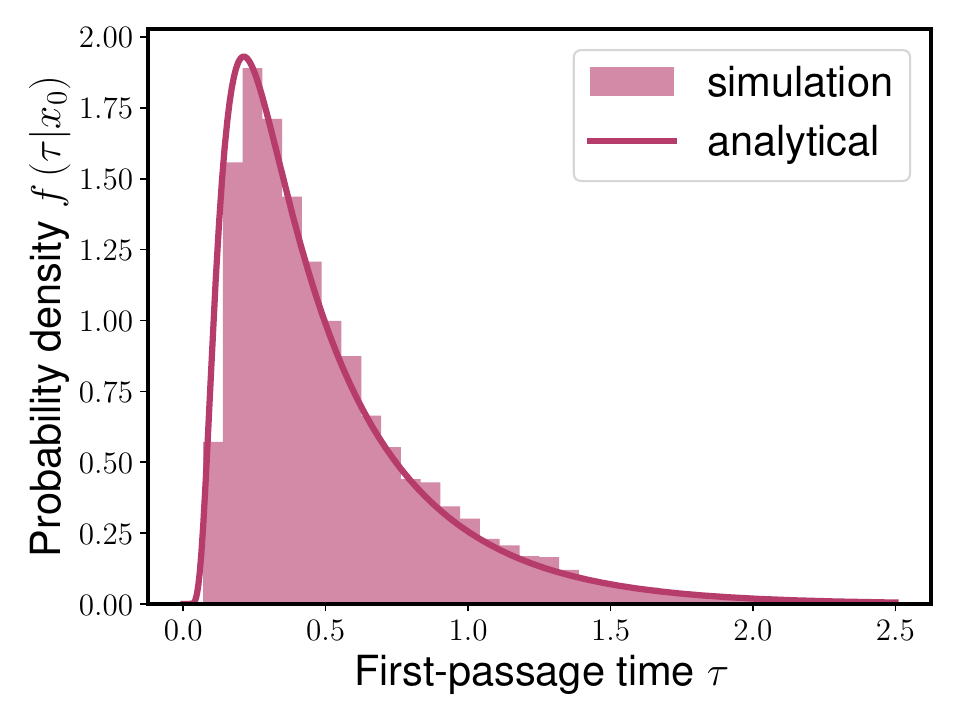}};
	\end{tikzpicture} 
        \caption{First-passage time distribution $ f(\tau \vert x_0)$ of leaving the interval $\plr{\varepsilon, 1-\varepsilon}$ at either end. a) Repeated quantum nondemolition measurements of a two-qubit system (Illustration 1 of the main text). b) Measurement-induced quantum synchronization of a five-qubit ring (Illustration 2 of the main text). Excellent agreement between the analytical prediction,  \cref{eq:fpt-distribution-full}, and the numerical simulation of the stochastic master equation \eqref{eq:diffusive-trajectory} is obtained. Same parameters as in the main text.}
        \label{fig:2}
    \end{figure}

 The first-passage time distributions for localization transitions of quantum trajectories can be directly derived
from the solution of the Fokker--Planck equation \eqref{eq:fpe-solution} \cite{kam97s,gar97s}. Assume the overlap at time $t=0$ is fixed.
The absorbing boundaries act like a drain and lead to a gradual decrease of the total probability still remaining in the interval $(\varepsilon,1-\varepsilon)$.
The probability that at time $t$ the trajectory is still surviving and has not yet left the interval is hence \cite{kam97s,gar97s}.
\begin{align}
  G(x_0,t) = \int_\varepsilon^{1-\varepsilon} \dd{x^\prime} p(x^\prime,t|x,0).
\end{align}
The rate of outflow thus identifies the first-passage time distribution with the probability of leaving the interval at time $\tau$
\begin{align}
  f(\tau|x_0) = -\dot{G}(x_0,\tau).
\end{align}
Using the Fokker--Planck equation \eqref{eq:fpe-solution} together with the definition of the survival probability, the first-passage time distribution of leaving the interval $\plr{\varepsilon, 1-\varepsilon}$ for trajectories starting at $x=x_0$ is then given by \cite{kam97s,gar97s}.
\begin{equation}
  f(\tau \vert x_0) 
  = -D(1-\varepsilon) \eval{\pdv{p(x^\prime,\tau)}{x^\prime}}_{x^\prime=1-\varepsilon}
  + D(\varepsilon) \eval{\pdv{p(x^\prime,\tau)}{x^\prime}}_{x^\prime=\varepsilon}.
  \label{eq:fpt-distribution-full}
\end{equation}
The one-sided first-passage time distributions of leaving the interval only to the left or only to the right are formally obtained from \cref{eq:fpt-distribution-full} by pushing the other absorbing boundary towards infinity so that it can never be reached.
Since the diffusion coefficient \eqref{eq:diffusion-coefficient} already makes the natural boundaries unattainable, in the present case, it is sufficient to extend the interval only up to $x=0$ or $x=1$.
The left- and right-sided first-passage time distributions are therefore
\begin{equation}
  f_1(\tau \vert x_0) = -D(1-\varepsilon) \eval{\pdv{p(x^\prime,\tau)}{x^\prime}}_{x^\prime=1-\varepsilon} \quad \text{and} \quad
  f_2(\tau \vert x_0) = D(\varepsilon) \eval{\pdv{p(x^\prime,\tau)}{x^\prime}}_{x^\prime=\varepsilon}.
  \label{eq:one-sided}
\end{equation}
Inserting $p(x,t)$ into the above formula for $f(\tau|x_0)$ \cref{eq:fpt-distribution-full} and using the symmetry properties 
\begin{align}
  D(x) = D(1-x), \quad F_{2n}^\prime(x) = F_{2n}^\prime(1-x), \quad F_{2n+1}^\prime(x) = -F_{2n+1}^\prime(1-x),
\end{align}
with $F_n^\prime(x) = \dv*{F_n(x)}{x}$, yields for the first-passage time distribution of leaving the interval
\begin{align}
  f(\tau \vert x_0) = \frac{1}{\gamma^2} D(x_0) D(\varepsilon)\sum_{n=1}^\infty F_{2n-1}(x_0) F_{2n-1}^\prime(\varepsilon) e^{-2\gamma^2 \lambda_{2n-1}\tau}
  \label{eq:fpt-distribution-full-2}.
\end{align}
The first-passage time distributions $f(\tau|x_0)$ for the illustrations of the main text are displayed in \cref{fig:2}.
Similarly, using \cref{eq:one-sided}, for the one-sided distributions it follows
\begin{align}
  f_1(\tau \vert x_0) &= -\frac{1}{2\gamma^2} D(x_0) D(1-\varepsilon) \sum_{n=1}^{\infty} F_{n}(x_0) F_{n}^\prime(1-\varepsilon) e^{-2\gamma^2 \lambda_{n}\tau}\label{eq:fpt-distribution-left-2},\\
  f_2(\tau \vert x_0) &= \frac{1}{2\gamma^2} D(x_0) D(\varepsilon) \sum_{n=1}^{\infty} F_{n}(x_0) F_{n}^\prime(\varepsilon) e^{-2\gamma^2 \lambda_{n}\tau}\label{eq:fpt-distribution-right-2}.
\end{align}

\section{Mean first-passage time and variance}
Using the time-homogeneity and employing the backward Fokker--Planck equation gives an evolution equation for the survival probability \cite{kam97s,gar97s}
\begin{align}
  \pdv{G(\tau,x_0)}{\tau} = \mu(x_0) \pdv{G(\tau,x_0)}{x_0} + D(x_0) \pdv[2]{G(\tau,x_0)}{x_0},
  \label{eq:G}
\end{align}
where $\mu(x)$ is a potentially state-dependent drift coefficient.
Applying integration by parts, the moments of the first-passage time distribution are found to be given by
\begin{align}
  \mathbb{E}[\tau^n](x_0) 
  = -\int_0^\infty \dd{\tau} \tau^n \pdv{G(\tau,x_0)}{\tau}
  = \int_0^\infty \dd{\tau} \tau^{n-1} G(\tau,x_0).
\end{align}
Together with \eqref{eq:G}, the moments can then be obtained recursively by means of the relation \cite{kam97s,gar97s}.

\begin{align}
  -n\mathbb{E}[\tau^{n-1}] = \mu(x)\pdv{x}\mathbb{E}[\tau^n] + D(x) \pdv[2]{x} \mathbb{E}[\tau^n].
\end{align}
Since the diffusion process we consider here is naturally drift-free, $\mu(x) = 0$, the mean first-passage time $\mathbb{E}[\tau](x_0)$ as a function of the initial overlap $x_0$ is determined by \cite{kam97s,gar97s}.
\begin{equation}
  -1 = D(x_0)\pdv[2]{\mathbb{E}[\tau](x_0)}{x_0}.
  \label{eq:mean-fpt-equation}
\end{equation}
Requiring general boundary conditions, $\mathbb{E}[\tau](a) = \mathbb{E}[\tau](b) = 0$, with $a,b\in\plr{0,1},\ a\le x_0 \le b,$ and integrating \cref{eq:mean-fpt-equation} twice yields
\begin{equation}
  \mathbb{E}[\tau](x_0) = \eta(x_0) - \frac{\eta(b)-\eta(a)}{b-a} x_0 + \frac{a\eta(b)-b\eta(a)}{b-a}
  \label{eq:mean-fpt}
\end{equation}
where
\begin{equation}
  \eta(x) = \frac{1}{\gamma^2}\plr{x-\frac{1}{2}}\ln(\frac{1}{x}-1).
\end{equation}
For $a=\varepsilon$ and $b=1-\varepsilon$, \cref{eq:mean-fpt} reduces to
\begin{equation}
  \mathbb{E}[\tau](x_0) = \eta(x_0) - \eta(\varepsilon).
  \label{eq:mean-fpt-reduced}
\end{equation}
The variance $\m{Var}(\tau\vert x_0) = \mathbb{E}[\tau^2](x_0) - \mathbb{E}[\tau]^2(x_0)$ can be similarly determined by integration of \cite{kam97s,gar97s}. 
\begin{align}
  D(x_0)\pdv[2]{x}\mathbb{E}[\tau^2](x_0) = -2\mathbb{E}[\tau],
\end{align}
with $\m{Var}({\tau\vert \varepsilon}) = \m{Var}({\tau\vert 1-\varepsilon}) = 0$.
We have
\begin{align}
  \text{Var}({\tau\vert x_0}) 
  = \frac{1}{\gamma^2}\mathbb{E}[\tau] + \eta^2({\varepsilon}) - \eta^2({x_0})
  - \frac{1}{4\gamma^4}\left[\ln(\frac{1}{x_0}-1)^2 - \ln(\frac{1}{\varepsilon}-1)^2\right],
\end{align}
with the same function $\eta(x)$ as in \cref{eq:mean-fpt}.
If the initial overlap has a given distribution $p(x_0) = p(x,0)$, then the moments of the first-passage time distribution follow as an average over $p(x,0)$ according to
\begin{align}
  \overline{\mathbb{E}[\tau^n]} = \int_\varepsilon^{1-\varepsilon} \dd{x_0} p(x_0) \mathbb{E}[\tau^n](x_0).
\end{align}

\end{document}